\documentclass[journal=jacsat,manuscript=article]{achemso}

\usepackage[version=3]{mhchem} 
\usepackage{tabularx}
\usepackage{booktabs}
\usepackage{multirow}
\usepackage{siunitx}
\usepackage{hyperref}
\usepackage[normalem]{ulem}
\usepackage[utf8]{inputenc}
\usepackage[dvipsnames]{xcolor}

\usepackage{xr} 

\author{Lorenzo Soprani}
\affiliation{Grenoble Alpes University, CNRS, Grenoble INP, Institut Néel, 38042 Grenoble, France}
\altaffiliation{Dipartimento di Chimica Industriale “Toso Montanari”, Università di Bologna, 40129 Bologna, Italy}
\author{Andrea Giunchi}
\affiliation{CINECA National Supercomputing Center, Casalecchio di Reno, I-40033 Bologna, Italy}
\altaffiliation{Dipartimento di Chimica Industriale “Toso Montanari”, Università di Bologna, 40129 Bologna, Italy}
\author{Marco Bardini}
\affiliation{Grenoble Alpes University, CNRS, Grenoble INP, Institut Néel, 38042 Grenoble, France}
\author{Quintin N. Meier}
\affiliation{Grenoble Alpes University, CNRS, Grenoble INP, Institut Néel, 38042 Grenoble, France}
\author{Gabriele D'Avino}
\affiliation{Grenoble Alpes University, CNRS, Grenoble INP, Institut Néel, 38042 Grenoble, France}
\altaffiliation{Department of Molecular Sciences and Nanosystems, Ca’ Foscari University of Venice, Venice, Italy}
\email{gabriele.davino@unive.it}

\title{Accurate and Efficient Phonon Calculations in Molecular Crystals via Minimal Molecular Displacements}

\abbreviations{DFT, PBE, DFTB, MMD, FC }

\begin{document}

\begin{tocentry}
\centering
\includegraphics[height=4cm]{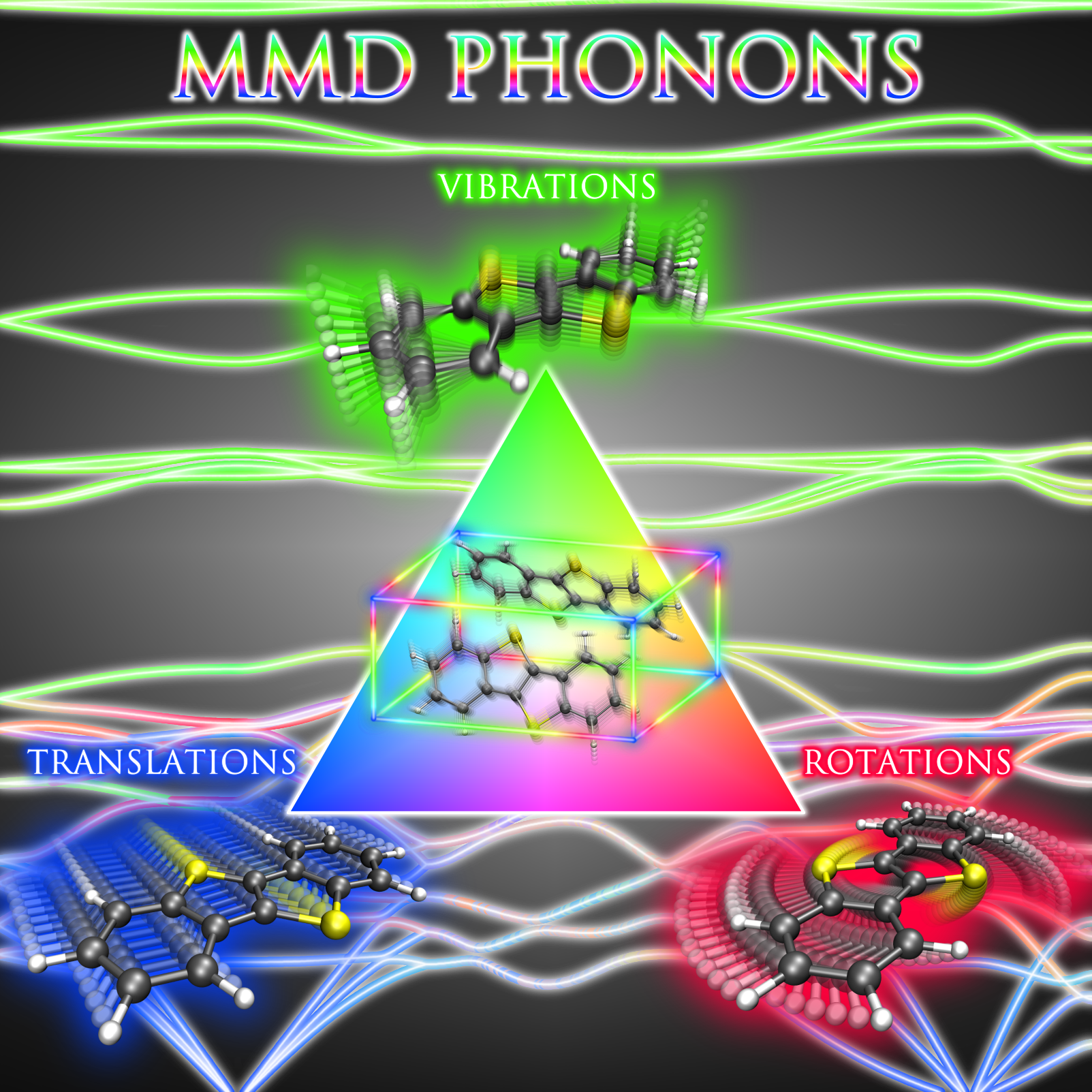}




\end{tocentry}

\clearpage
\begin{abstract}
Vibrational dynamics governs the fundamental properties of molecular crystals, shaping their thermodynamics, mechanics, spectroscopy, and transport phenomena.
However desirable, the first-principles calculation of solid-state vibrations, i.e.\ phonons, stands as a major computational challenge in molecular crystals characterized by many atoms in the unit cell and by weak intermolecular interactions. 
Here we propose a formulation of the harmonic lattice dynamics based on a natural basis of molecular coordinates consisting of rigid-body displacements and intramolecular vibrations. 
This enables a sensible \emph{minimal molecular displacement} approximation for the calculation of the dynamical matrix, combining isolated molecule calculations with only a small number of expensive crystal supercell calculations, ultimately reducing the computational cost by up to a factor 10.
The comparison with reference calculations demonstrates the quantitative accuracy of our method, especially for the challenging and dispersive low-frequency region it is designed for.
Our method provides an excellent description of the thermodynamic properties and offers a privileged molecular-level insight into the complex phonons band structure of molecular materials.
\end{abstract}


\clearpage
\section{Introduction}
\label{s:intro}

Vibrational dynamics plays a central role in the physics and chemistry of molecular crystals, as it governs their thermodynamics, mechanical properties, spectroscopic behavior, charge and heat transport phenomena, among others. As such, the vibrations of molecular solids are relevant to several domains of high practical relevance, spanning from pharmaceutical industry to functional materials.
The vibrational dynamics of molecular crystals is the subject of an intense experimental and theoretical research  effort in the context of organic semiconductors,\cite{Coropceanu09,Girlando10,Chernyshov17,Tu18,Xie18,Harrelson19,Ruggiero19,Banks22,Banks23} whose development suffer for the low charge carriers’ mobility that rarely exceeds the value of \SI{10}{\square\cm\per\V\per\s}.\cite{Fratini20,Schweicher20}
Charge transport in van der Waals molecular solids is strongly hampered by the effect of large-amplitude molecular motion associated with low-frequency lattice vibrations.\cite{Eggeman13,Illig16} 
In fact, in these soft materials the energetic disorder due to the thermal lattice motion is comparable to the electronic bandwidth, leading to charge carriers that transiently localize over distances comparable to the lattice spacing.\cite{Fratini16,Giannini23} 
The reliable and efficient calculation of lattice vibrations for the identification and the possible suppression of detrimental mobility \emph{killer modes} stands as an important challenge for the improvement of the transport properties of organic materials.\cite{Schweicher19,Stockel21,Dettmann23}

The calculation of harmonic phonons in molecular crystals follows the general and established framework for crystalline systems.
The main established method for calculating phonon frequencies from first principles is density function theory (DFT), the two main computational routes being density functional perturbation theory\cite{Gonze97,Baroni01} and the frozen-phonon method, i.e.\ finite differences of analytic forces developing upon atomic displacements from equilibrium.\cite{Togo_2023}
While the former method, based on analytical derivatives, is more efficient for relatively small systems, frozen phonon is usually the technique chosen for molecular crystals, also thanks to the possibility of an embarrassingly parallel workload.
Nevertheless, the frozen-phonon DFT calculation of the lattice dynamics in molecular crystals is a computationally challenging and demanding task.
This is the result of two factors.
First, weak intermolecular interactions require a very high numerical accuracy for the reliable calculation of the dynamical matrix, as displacements from equilibrium result in tiny variations in energy and forces.
This calls for very stringent numerical settings.
Second, molecular crystals typically feature large unit cells, often containing over one hundred atoms, which becomes even more problematic in supercell calculations needed to obtain phonon dispersion.

The combination of the two factors mentioned above severely limits the feasibility of phonon calculations, making cheaper alternatives highly desirable.
Classical force fields have been uses with a certain degree of success for oligoacenes, \cite{Coropceanu09,Sanchez10,Venuti02,DellaValle08} albeit the availability and reliability of parameters for heteroatoms or specific functional groups remains an important limitation.
Different parameterizations of semiempirical density functional tight binding methods (DFTB) have been employed in this context\cite{Xie18,Dettmann21}.
The benchmark work by Kamencek \emph{et al.}, highlighted systematic flaws of DFTB on the description of the unit cell volume and the  vibrational spectrum, even for a system as simple as naphthalene.\cite{Kamencek20}
Machine learning potentials trained on DFT data offer a promising route for accurate atomistic potentials in molecular crystals,\cite{Musil18,Zugec24} which however calls for huge database of reference calculations to explore the overwhelmingly vast chemical space, and whose accuracy for the highly-dispersive lattice phonon region is yet to be demonstrated. 
Ultimately, when precision is required, DFT with appropriate dispersion corrections \cite{Bedoya18} remains the sole viable, however costly, option.

Although phonon calculations in molecular crystals do not fundamentally differ from those in other solids, it is important to highlight certain characteristics that are common to the broader class of molecular crystals.
These features will be essential for the present development and will enable major computational savings with a minimal impact on accuracy.
The vibrational properties of molecular crystals lays in between those of individual molecules and covalent solids.\cite{Califano81} 
Indeed, these solids can be seen as ordered assemblies of molecular units, each of which is characterized by strong chemical bonds between atoms, and much weaker non-covalent interactions (e.g.\ van der Waals, electrostatics) between molecules.
This results in distinct energy and time scales for the resulting vibrations, with intermolecular lattice modes appearing in the low-frequency region (usually below $\SI{200}{\per\cm} \approx \SI{6}{\tera\hertz}$, also referred to as thermal phonons) and intramolecular ones at higher energy.
As a matter of fact, the vibrational spectra (Raman or infrared) of molecular crystals are virtually identical to their molecular analogues at high frequency (except for selection rules imposed by the  crystal symmetry), while the THz domain is highly sensitive to the fine details of the crystal packing.
THz spectra have been proposed as a reliable fingerprint for the practical and non-destructive identification of crystal polymorphs. \cite{Venuti02,Salzillo18,Giunchi19}
Moreover, THz modes are strongly dispersive, while higher-frequency vibrations typically present a small to negligible wave vector dependence,\cite{Natkaniec80} further attesting their intramolecular nature.

The boundary between molecular and lattice vibrations is not, in general, a sharp one, as those modes do inevitably mix, especially when they are characterized by comparable energies.
Girlando and collaborators \cite{Girlando10} set up an hybrid computational scheme to compute Brillouin-zone center ($\Gamma$ point) phonons by combining DFT calculations on isolated molecules with model potentials for intermolecular interactions.
This effective approach to mode mixing provided a reliable description of lattice modes that was then used to compute electron-phonon coupling in the prototypical organic semiconductors rubrene and pentacene. \cite{Girlando10,Girlando11}
By computing phonon band structures for different oligoacene crystals, Kamencek and Zojer showed that a substantial mixing between rigid-body molecular motions (i.e.\ translation and rotations) and intramolecular deformations occurs in larger molecules, such as tetracene and pentacene. \cite{Kamencek22}

In this paper we present a novel approach to the harmonic lattice dynamics of molecular crystals.
This consists in a frozen-phonon method built on the basis of molecular displacements (rigid translations and rotations, plus isolated-molecule normal modes), instead of the usual set of atomic displacements.\cite{Califano81}
For a complete set of coordinates, our method is equivalent to a conventional frozen phonon calculation.
The key advantage of the molecular route is that it sets the basis for the development of a sensible approximation to the phonon problem, termed the minimal displacement (MMD) method, which allows for a four- to ten-fold reduction of the computation time. 
The capabilities of this approach, coupled to highly-accurate plane-wave DFT calculations, are here explored for a benchmark set of typical molecular crystals of interest for organic electronics applications.
We demonstrate that the MMD approximation implies a negligible accuracy loss for thermal phonons, achieving a globally very good accuracy also at higher frequencies.

This paper is organized as follows.
The next section presents the methodology, first setting up a general framework for the description of lattice dynamics in terms of molecular displacement coordinates, and then introducing the MMD approximation. 
The result section illustrates the application of our methodology and quantifies the accuracy on phonon frequencies achievable within the MMD scheme.
The presentation starts from Brillouin-zone center ($\Gamma$ point) results, to then encompass full lattice dynamics calculations including phonon dispersion and the resulting vibrational contribution to thermodynamic properties. 
The results section also addresses the computational savings achieved with the MMD scheme.
The main points and perspectives of this work are discussed in the closing section.

\section{Methodology}
\label{s:methods}

\subsection{The molecular displacement basis}
The core of any phonon calculation consists in the computation of the force constant (FC) matrix 
\begin{equation}
\label{e:FCmat}
\Phi_{ij}^\mathrm{(A)}= \frac{\partial^2 E}{\partial x_i \partial x_j} =
\frac{\partial F_i}{\partial x_j}
\end{equation}
composed of the second derivatives of the total energy $E$ with respect to atomic displacements from equilibrium $x_i$ or, equivalently, on the first derivatives of the forces $F_i$.
The frozen-phonon method evaluates the FC matrix by finite differences of the forces by performing small displacements.
The displacements $x_i$ usually correspond to the $3N$ Cartesian coordinates of the atoms in the unit cell or in a supercell, from which the superscript A in $\Phi_{ij}^\mathrm{(A)}$.

\begin{figure}
\centering
\includegraphics[width=.6\textwidth]{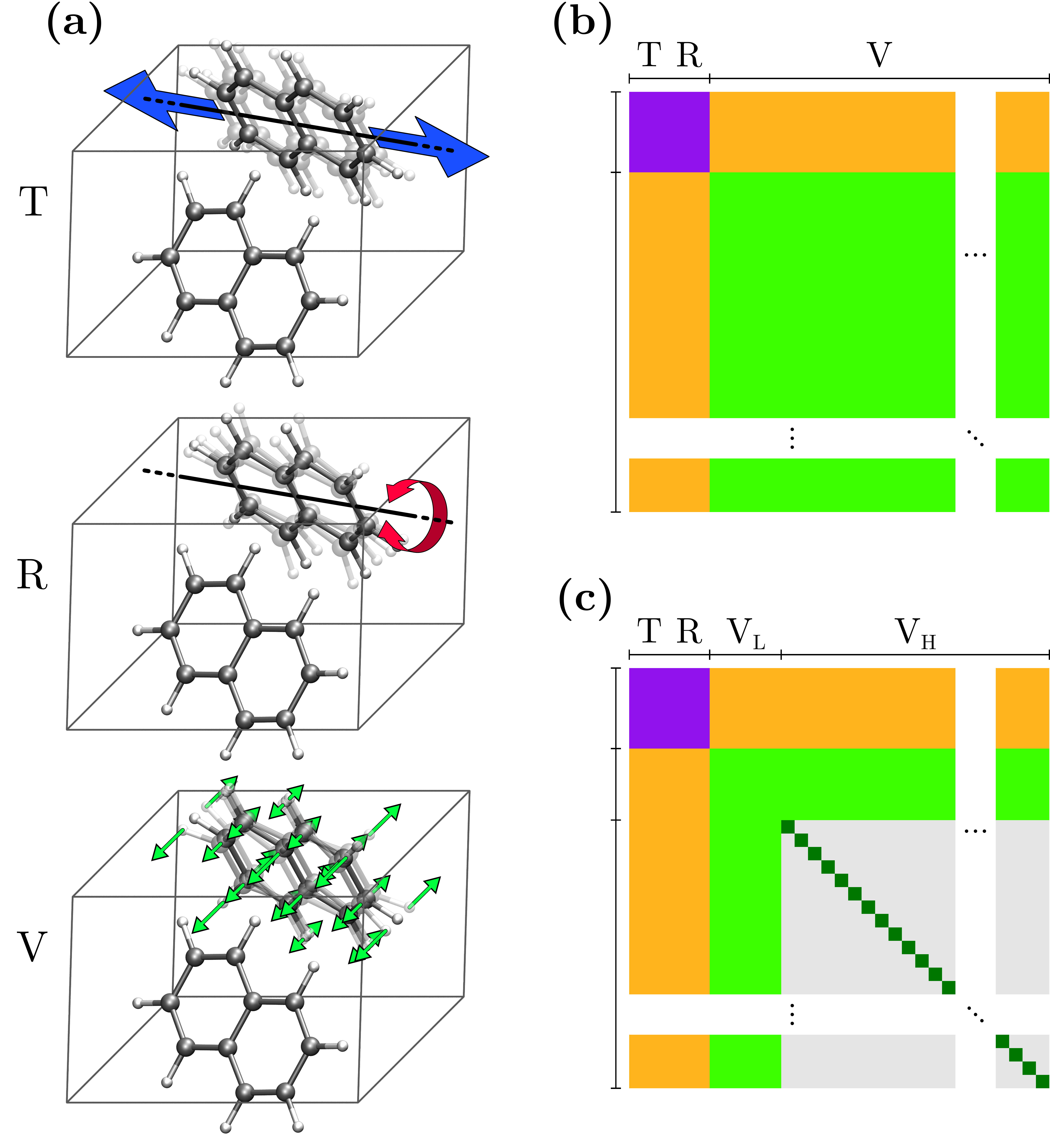}
\caption{(a) Illustration, based on the naphthalene crystal, of a T, R and V molecular displacements employed in the calculation of the force constant matrix.
Structure of the force constant matrix in the basis of molecular displacements for (a) an exact calculation and (b) according to the MMD approximation.
Purple and green regions respectively correspond to the blocks of rigid molecular motion and isolated-molecule distortions, the orange area being the coupling between the two.
In the MMD framework, the off-diagonal elements of the V$_\textrm{H}$ block are set to zero (light gray area), and the diagonal ones taken from the normal modes of the isolated molecule (in dark green).}
\label{f:FCmatrix}
\end{figure}

Building on the molecular nature of the system, we define an alternative set of molecular-displacement coordinates as a linear combinations of Cartesian atomic displacements
\begin{equation}
\label{e:moldisp}
u_j=\sum_i U_{ij} x_i .
\end{equation}
For each molecule we introduce a set of rigid-body motions, corresponding to the three translations (T) and the three rotations (R) that are both conveniently referred to the molecular inertial frame.
In accordance with the harmonic approximation, rotations are linearized to comply with Equation~\ref{e:moldisp}.
We complement the rigid-body displacements with the normal modes of the isolated molecule (V), in which translations and rotations have been projected out by applying the Eckart conditions.
This operation can be iterated for each molecule in the (super)cell, fully defining the change-of-basis matrix $U$.
For illustrative purpose, Figure~\ref{f:FCmatrix}a displays the T along the long molecular axis, the R about the same axis and an intramolecular V displacement (see also Supporting Information, SI, Figure~\ref{SI-f:disp}).

The new set of molecular displacements in Eq.~\ref{e:moldisp} and the full set of atomic displacements are two complete bases that can be equivalently used to set up the vibrational problem in the harmonic approximation.
We can hence write the FC matrix in this new basis without loss of generality
\begin{equation}
\label{e:FCmat_M}
\Phi_{ij}^\mathrm{(M)}= \frac{\partial^2 E}{\partial u_i \partial u_j}=
\frac{\partial F^\mathrm{(M)}_i}{\partial u_j},
\end{equation}
where $F^\mathrm{(M)}_i= \sum_k U_{ki}\, F_k $ is the projection of the vector of the forces onto the molecular displacement $u_i$.
The FC matrices expressed in the atomic and in the molecular displacements coordinates are related by the basis-change transformation
\begin{equation}
\label{e:FC_basis_change}
{\Phi}^\mathrm{(M)} = U^{T}\,  \Phi^\mathrm{(A)} \, U.
\end{equation}
In both cases, one should evaluate the forces on all the atoms in the (super)cell for every  displacement, or twice as much for central finite differences.
The structure of the FC matrix in the molecular-displacements basis, is sketched in Figure~\ref{f:FCmatrix}b, where the purple and green regions represent the block of rigid molecular motions (T and R modes) and intramolecular vibrations (V modes), respectively.
The orange areas depict the coupling between the two subspaces.

\subsection{The minimal molecular displacement approximation}
\label{s:mmmd}

Working with molecular displacements offers an advantage for anyone interested in a specific spectral window, such as the challenging and highly dispersive low-frequency region of room-temperature thermal vibrations.
The advantage consists in the possibility of computing an approximated FC matrix by considering only a small subset of molecular displacements (i.e.\ rigid-body motions plus a few low-frequency V modes) for which forces are actually computed with expensive solid-state DFT calculations.
Such an approximated FC matrix is then complemented with isolated-molecule vibrational data for high-frequency V modes, resulting in a minimal molecular displacement (MMD) approximation to the solid-state vibrational problem.
As we will later show, the MMD approximation preserves most of the accuracy of a full phonon calculation at a fraction of the computational cost.
To set up the MMD scheme, intramolecular V displacements are split into a low- and a high-frequency subset (V$_\textrm{L}$ and V$_\textrm{H}$) by comparing the frequencies obtained from the normal-mode analysis on the isolated molecule against a cutoff frequency $\tilde \nu_\mathrm{cut}$.
As we shall see in the following, 
$\tilde{\nu}_\mathrm{cut}\approx \SI{200}{\per\cm}$ is a plausible threshold value.
The V$_\textrm{L}$ block contains the modes that will significantly mix with rigid-body motions, which are a small fraction (5-\SI{15}{\percent}) of the total number of V modes.

The structure of the FC matrix obtained within the MMD approximation is illustrated in Figure~\ref{f:FCmatrix}c.
The purple, green, and orange regions are identical to those of the full FC matrix in panel a.
The approximation instead involves the V$_\textrm{H}$-V$_\textrm{H}$ block, with matrix elements $\Phi_{ij} = \Phi^\mathrm{mol}_{i} \delta_{ij}$, where $\Phi_i^\mathrm{mol}$ is the FC of the normal mode $i$
of the isolated molecule.
The MMD approximation therefore consists in neglecting the direct coupling among high-frequency intermolecular displacements (light gray region in Figure~\ref{f:FCmatrix}c) and in adopting the FCs of the isolated molecule as the diagonal elements of $\Phi^\mathrm{(M)}$ (dark green elements).
We note that even in the absence of a direct interaction within the molecular coordinates of the V$_\textrm{H}$ block, high-frequency displacements do nevertheless couple to T, R and V$_\textrm{L}$ modes and among themselves (indirectly, though the coupling with T, R and V$_\textrm{L}$ modes).

\subsection{Phonon dispersion in the MMD approximation}
\label{s:general_disp}

Having constructed the dynamical matrix within the molecular displacement basis, and possibly within the MMD approximation, one can obtain the analogue in the basis of atomic displacement upon basis-change transformation, i.e.\ ${\Phi}^\mathrm{(A)} = U^{-\mathrm{T}}\, \Phi^\mathrm{(M)}\, U^{-1}$.
The calculation of the harmonic lattice dynamics can then proceed according to the ordinary procedure for Cartesian atomic displacements.
Phonon frequencies ($\omega_{\lambda}$) and eigenvectors ($\bar \epsilon_{\lambda}$)
can be obtained for each $\mathbf{q}$ point by solving the eigenvalue problem
\begin{equation}
\label{e:eigval}
D^\mathrm{(A)}(\mathbf{q}) \bar \epsilon_{\lambda}(\mathbf{q}) = \omega_{\lambda}^2(\mathbf{q}) \ \bar \epsilon_{\lambda}(\mathbf{q}) 
\end{equation}
where we have introduced the dynamical matrix with elements
\begin{equation}
\label{e:dynmat_A}
D^\mathrm{(A)}_{jk}(\mathbf{q})=\frac{1}{\sqrt{M_j M_k}} 
\sum_\mathbf{t} \Phi_{\mathbf{t} j, \mathbf{0} k} 
\ e^{-i \mathbf{q} \cdot\left(\mathbf{t} +\mathbf{r}_j-\mathbf{r}_k\right)} \ .
\end{equation}
The indexes $j$ and $k$ run on the $3N$ Cartesian components of the atoms in the unit cell, $M_j$ ($\mathbf{r}_j$) the mass (vector position) of atom of coordinate $j$, $i$ is the imaginary unit, and the translation vector $\mathbf{t}$ runs on the unit cells of the crystal;
$\Phi_{\mathbf{t} j, \mathbf{0} k}$ is the FC matrix element between the coordinate $j$ of an atom in cell $\mathbf{t}$ and the coordinate $k$ of an atom in the cell at the origin.
Practical calculations make use of finite supercells, leading to FC matrix truncated in the real space (Fourier interpolation) at $\mathbf{q}$ points not commensurate with the chosen supercell.

Alternatively, one can formulate the eigenvalue problem in Equation~\ref{e:eigval} on the basis of molecular displacements.
While the discrete Fourier transform in Equation~\ref{e:dynmat_A} is naturally performed in the Cartesian atomic basis, the dynamical matrix can be transformed to the molecular displacement basis as
\begin{equation}
D^\mathrm{(M)}(\mathbf{q})= 
 \widetilde{U}^\mathrm{T}  D^\mathrm{(A)}(\mathbf{q}) \, \widetilde{U}
\label{eq:dynmat_M}
\end{equation}
making use of the unitary change-of-basis matrix
\begin{equation}
    \widetilde{U}_{ij} =
    \frac{M_{i} \, U_{ij}}{\sqrt{ \sum_k \left( M_{k} \, U_{kj} \right)^2}} \ .
\label{eq:Ptilde}
\end{equation}
In the present work, the latter formulation in terms of molecular displacements is adopted.
This permits retaining the information on the molecular nature of the system, bringing direct insight on the crystal lattice dynamics.

\subsection{Computational details}
\label{s:details}

Frozen phonon calculations in the atomic and molecular displacement basis have been implemented in the in-house code \textsf{PhonoMol} that provides a unique interface to molecular and solid-state DFT engines, performing the pre- and post-processing operations. 
\textsf{PhonoMol}, written in Python 3, works on the primitive unit cell, making use of the \textsf{Spglib} library\cite{spglib} for the standardization of experimental crystal structures and the handling of the space group symmetry.

Our implementation relies on central finite differences for the evaluation of the derivatives of forces on atoms with respect to 
atomic or molecular displacements (Equations~\ref{e:FCmat},\ref{e:FCmat_M}). 
A displacement amplitude of \SI{0.005}{\angstrom} has been adopted throughout this work. 
For molecular displacements, this value corresponds to the magnitude of the maximum atomic displacement realized by the collective molecular motion.
Translational and point symmetry are fully exploited in order to minimize the number of solid-state DFT calculations to be performed for displaced geometries.
FC matrices have been computed from supercell calculations, using a $2\times 2\times 2$ replica of the unit cell for all systems.
Phonon densities of states (DOS) and thermodynamic properties have been computed using a uniform $8\times 8\times 8$ sampling of the Brillouin zone, ensuring converged results.
The DOSs have been computed as a sum of Gaussian peaks with standard deviation of 2.5~cm$^{-1}$.

The proposed method is generally applicable to any atomistic potential, ranging from different \emph{ab initio} flavors to force fields.
Our implementation relies on solid-state DFT calculations with dispersion corrections.
Specifically, we employ the Perdew–Burke-Ernzerhof (PBE) functional\cite{PBE} along with the projector-augmented wave method (PAW, version 6.4)\cite{Blochl94,Kresse99} and Grimme D3 dispersion corrections with Becke-Johnson damping (D3-BJ).\cite{Grimme10,Grimme11}
The Vienna Ab initio Simulation Package (VASP, {version 6.4.0}),\cite{VASP1,VASP2,VASP3,VASP4} has been used for DFT calculations.
Numerical settings have been carefully chosen in order to ensure the level of accuracy needed for phonon calculations in soft molecular crystal, ensuring the convergence of the total energy within \SI{1}{\milli\eV}/atom.
These include the Brilluoin zone sampling mesh that has been optimized for each system (see SI Table~\ref{SI-t:KPOINTS}), the plane wave energy cutoff set to \SI{800}{\eV}, the self-consistent field (SCF) convergence criterion set to \SI{e-9}{\eV}.
The atomic position have been relaxed at unit cell parameters fixed to the experimental values until reaching a residual force below \SI{e-3}{\eV\per\angstrom}.
This methodology proved able to achieve an excellent agreement with THz Raman spectroscopy data on molecular crystals.\cite{Bedoya18,Salzillo18}

Isolated molecule DFT calculation have been performed with the ORCA ({version 5.0}) package.\cite{ORCA,ORCA5}
For consistency with solid-state calculations, we employed the PBE functional\cite{PBE} with D3-BJ dispersion corrections\cite{Grimme10,Grimme11} and the def2-SVP Gaussian atomic orbitals basis.\cite{Weigend05} 
Tight convergence criteria have been adopted in geometry optimization and frequency calculations.

\section{Results}
\label{s:results}

\begin{figure}
\centering
\includegraphics[width=0.67\linewidth]{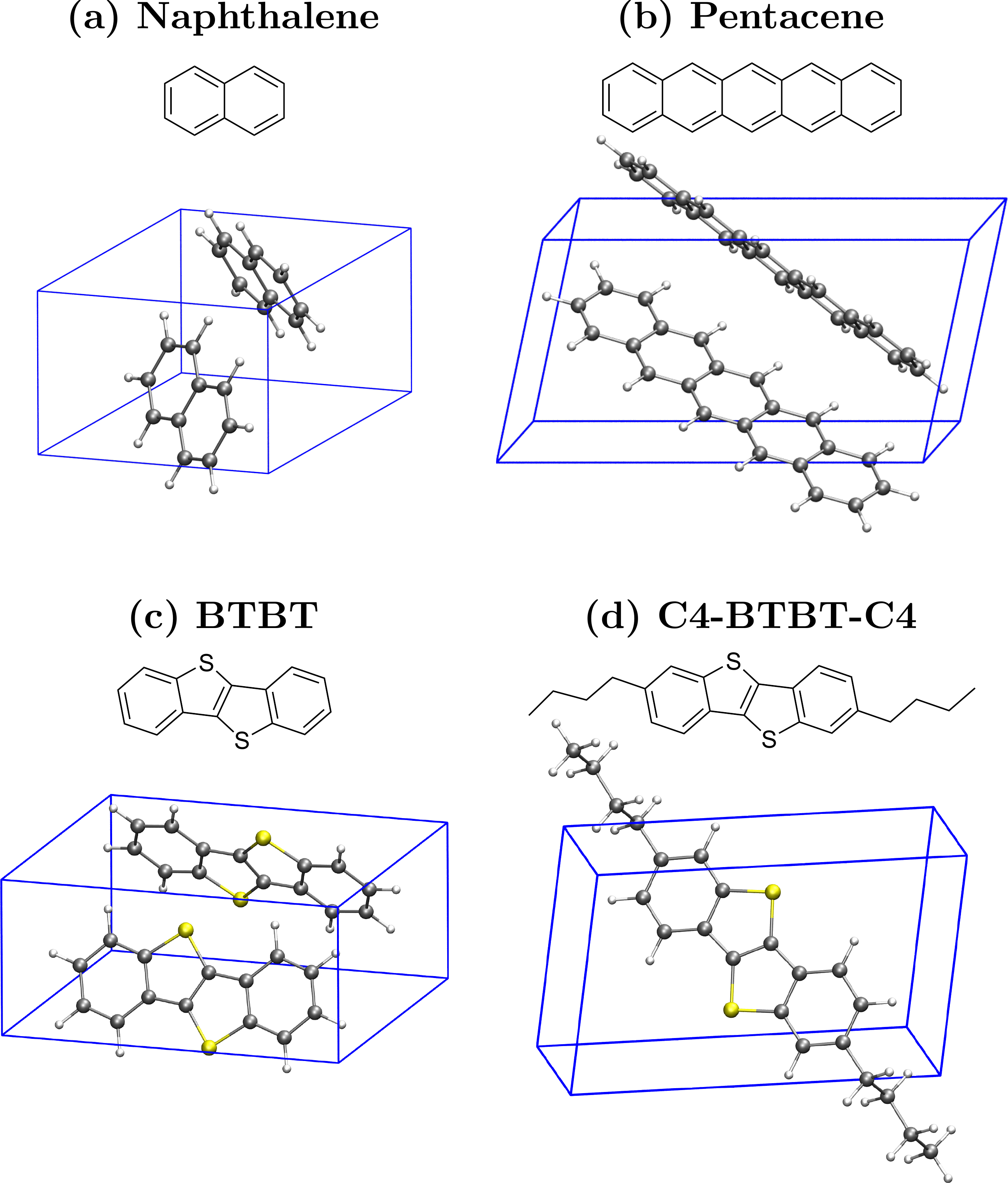}
\caption{Molecular and crystal structure of the four compounds considered in this work.}
\label{f:mols}
\end{figure}

With the methodology established, we now present the phonon calculations for four conjugated molecular crystals relevant to organic electronics -- see Figure~\ref{f:mols}.
The set includes two oligoacenes,  naphthalene and pentacene, as well two thienoacenes compounds, namely BTBT ([1]benzothieno[3,2-\textit{b}][1]benzothiophene) and its alkylated derivative C4-BTBT-C4. 
The latter serves as a benchmark system for molecular semiconductors functionalized with floppy side chains, which are often grafted to the functional conjugated core to improve processability.
The crystal structures for the four compounds have been taken from the Cambridge Crystallographic Data Centre, with codes: 600182,\cite{Capelli06} 170187,\cite{Mattheus01} 2166518,\cite{Akai22} 1525675.\cite{Minemawari17}
See SI Table~\ref{SI-t:cell-prms} for unit cell parameters.

In the following, we will compare reference phonon calculations obtained by constructing the FC matrix with a complete set of displacements with the results obtained with the MMD approximation.
While recalling that reference calculations obtained with atomic and molecular displacements yield the same results within numerical accuracy, we note that MMD results depend on the choice of the cutoff frequency (see Methods).
In order to assess the impact of this critical parameter on the quality of the MMD method, we will present results obtained for two different cutoffs, namely 
$\tilde \nu_\mathrm{cut}^{(1)}\approx$~\SI{200}{\per\cm} and 
$\tilde \nu_\mathrm{cut}^{(2)}\approx$~\SI{400}{\per\cm}.
The actual values, which vary from system to system, are given in Table~\ref{t:cutoff} and correspond to the highest-frequency isolated-molecule vibration that is employed for computing the FC matrix in solid-state calculations.
The number of isolated-molecules modes with wavenumber $\le\tilde \nu_\mathrm{cut}$, $N_{\textrm{VL}}$, is also reported in Table~\ref{t:cutoff}.
As a rule of thumb, we recommend choosing the cutoff frequency after the inspection of the vibrational spectrum of the isolated molecule, e.g.\ by avoiding to cut through manifolds of modes very close in frequency, but rather placing the threshold in a gap. 
We remark that phonon frequencies are stable against small variations of $\tilde \nu_\mathrm{cut}$, although the inclusion or exclusion of one or a few V displacements might affect the Fourier interpolation of the first acoustic branch.

\begin{table}[]
\caption{Definition of the system-specific cutoff wavenumbers $\tilde \nu_\mathrm{cut}^{(1)}$ and $\tilde \nu_\mathrm{cut}^{(2)}$ (in \unit{\per\cm}) and of the number of isolated-molecule vibrations with frequencies less than or equal to each threshold, $N_\mathrm{VL}$.
The latter corresponds to the number of V displacements that will be considered in the computation of the force constant matrix in the MMD method.}
\centering
\begin{tabular}{lS[table-format=3.0]S[table-format=2.0]S[table-format=3.0]S[table-format=2.0]}
\toprule
{System} & {$\tilde \nu_\mathrm{cut}^{(1)}$} & 
{$N_\mathrm{VL}^{(1)}$} & {$\tilde \nu_\mathrm{cut}^{(2)}$} & {$N_\mathrm{VL}^{(2)}$}  \\
\midrule
Naphthalene & 182 &  2 & 386 &  4 \\
Pentacene   & 238 &  8 & 373 & 13 \\ 
BTBT        & 235 &  6 & 368 &  9 \\
C4-BTBT-C4  & 204 & 15 & 396 & 25 \\
\bottomrule
\end{tabular}
\label{t:cutoff}
\end{table}

\subsection{Brillouin-zone center modes}

\begin{figure}
\includegraphics[width=\textwidth]{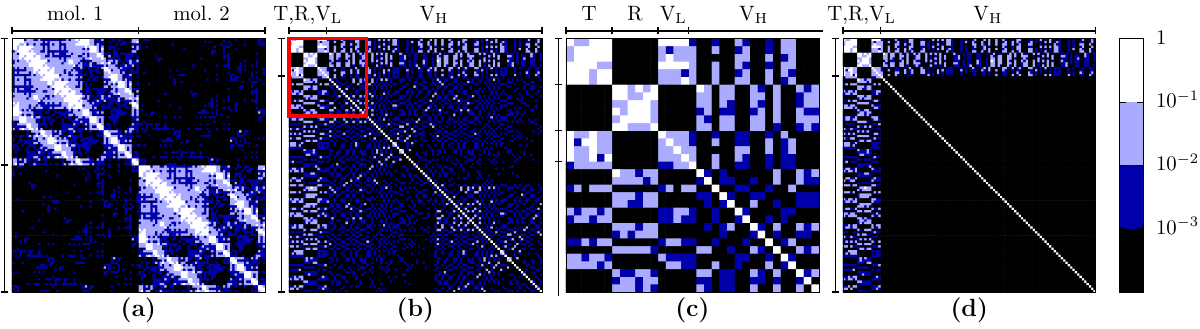}
\caption{Normalized $\Gamma$-point dynamical matrix for the naphthalene crystal, expressed (a)~in the full basis of atomic Cartesian displacements and (b) in the full basis of molecular displacements, including translations (T), rotations (R) and intramolecular modes (V).
(c) Zoom of the region highlighted with a red frame in the previous panel.
(d) Normalized dynamical matrix obtained in the MMD approximated scheme, with off-diagonal elements coupling high-frequency molecular modes (V$_\textrm{H}$) set to zero.
This can be obtained at a reduced computational cost by performing gradient calculations only for rigid-molecule displacements (T,R), plus a small number of low-frequency molecular modes (V$_\textrm{L}$).}
\label{f:dynmat}
\end{figure}

We start the presentation by comparing the dynamical matrix at $\Gamma$ point, expressed in the usual Cartesian atomic displacement basis with its analogue obtained with molecular displacements.
This is shown in Figure \ref{f:dynmat} for the illustrative case of the naphthalene crystal, including two molecules in the unit cell.
In order to convey information on the relative magnitude of off-diagonal elements as compared to the diagonal elements of the coupled displacements, the figure shows the normalized dynamical matrix
$\bar D_{ij}= |D_{ij}|/\sqrt{D_{ii} D_{jj}}$.

Figure \ref{f:dynmat}a shows $\bar D$ expressed in the basis of the Cartesian atomic displacement.
When atomic displacements are grouped molecule-wise, two diagonal blocks corresponding to the two molecules in the unit cell can be easily identified.
Conversely, the two off-diagonal blocks describe the couplings between displacements of atoms belonging to different molecules.
The matrix elements of the diagonal blocks are typically two to three orders of magnitude larger than those of the off-diagonal blocks, due to the fact that the forces developing on a given atom are much larger when atoms of the same molecule are displaced, as compared to the effect of a displacement of an atom of another molecule.
This reflects the fact that atoms belonging to the same molecules are bound by strong covalent interactions, while atoms of different molecules interact through much weaker non-covalent forces.

The normalized dynamical matrix radically changes upon switching to the molecular displacements basis, see Figure~\ref{f:dynmat}b.
It is convenient to sort these new displacements in the following order: translations (T), rotations (R) and isolated-molecule vibrations (V), the latter ranked in ascending frequency order. 
This leads to the matrix structure characterized by a checkerboard pattern in the upper-left corner, that corresponds to the block of T, R and low-frequency intramolecular modes, V$_\textrm{L}$ (see also zoom of the red frame in panel c).
Indeed, as far as $\Gamma$ phonons are concerned, R and T do not mix as they belong to different irreducible representations of the symmetry group, consistent with the group factor analysis from rigid molecules.\cite{Turrell72}
We further note that the off-diagonal matrix elements in the T-R-V$_\textrm{L}$ subspace have the largest magnitude, making these modes more susceptible to mixing.
Moreover, the off-diagonal elements coupling T-R-V$_\textrm{L}$ modes with high-frequency intramolecular vibrations (V$_\textrm{H}$) are, on-average, larger than those of the V$_\textrm{H}$-V$_\textrm{H}$ block.
These observations offers strong indication of the suitability of the MMD approximation.

Figure~\ref{f:dynmat}d shows the normalized dynamical matrix $\bar D$ that would be obtained within the MMD approximation, i.e.\ with all the off-diagonal matrix elements of the V$_\textrm{H}$-V$_\textrm{H}$ block set to zero.
We recall that the computation of $\bar D$ requires a minimal number of periodic DFT calculations corresponding to T, R and V$_\textrm{L}$ modes.
As such, this can be obtained at a small fraction of the computational cost of the full dynamical matrix.

\begin{figure}
\includegraphics[width=1\textwidth]{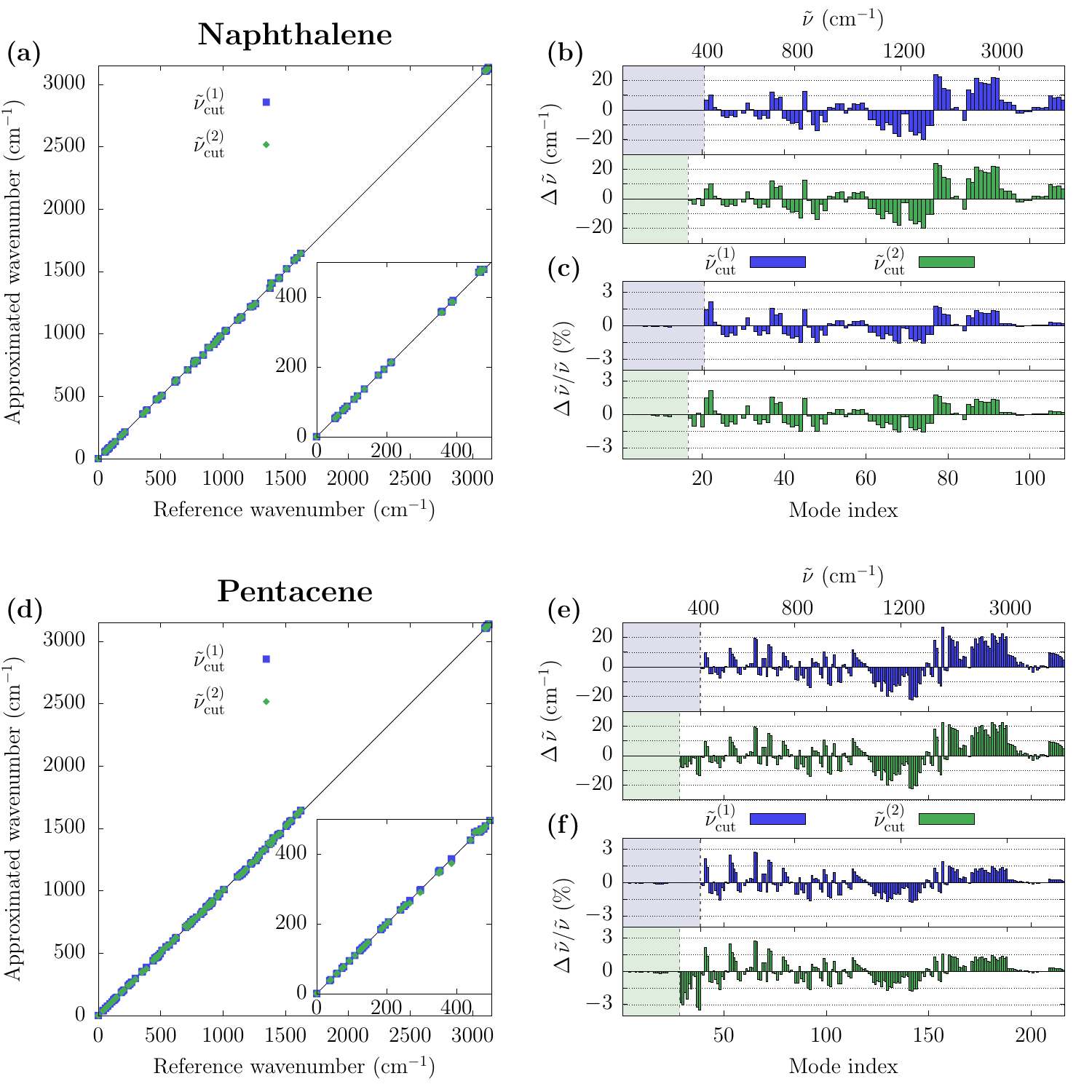}
\caption{Comparison between the exact and MMD-approximated phonon frequencies for the naphthalene (panel a-c) and pentacene (panels d-f) crystals.
MMD results refer to approximated dynamical matrices built with $\tilde{\nu}_\textrm{cut}^{(1)} \approx \SI{200}{\per\cm}$ and $\tilde{\nu}_\textrm{cut}^{(2)} \approx \SI{400}{\per\cm}$ cutoff criterion for the selection of V$_\textrm{L}$ modes (see Table~\ref{t:cutoff} for exact cutoff values).
(a,d) Plots of the exact vs.\ approximated frequencies.
The insets provide a zoom of the low-frequency region.
(b,e) Difference and (c,f) relative difference between MMD and reference vibrational frequencies.
Values are reported against the mode index, ranked in ascending frequency order.
An approximate top x-scale reporting mode frequencies is provided as an indication.}
\label{f:freq_NAP-PEN}
\end{figure}

We are now ready to quantify the accuracy of the proposed approximation, taking the $\Gamma$-point phonon frequencies of naphthalene and pentacene as a first benchmark.
Figure~\ref{f:freq_NAP-PEN}  compares reference results with MMD ones obtained for two different values of the cutoff frequency, 
$\tilde{\nu}_\textrm{cut}^{(1)}$ and $\tilde{\nu}_\textrm{cut}^{(2)}$ -- similar data for the other compounds are shown in SI Figure~\ref{SI-f:freq-BTBT-C4BTBTC4}.
For both naphthalene and pentacene, the MMD-approximated calculations yield vibrational frequencies that are in very good agreement with reference calculations, as can be appreciated in the first instance from the nearly ideal linear plots of exact vs.\ approximated frequencies in \figurename~\ref{f:freq_NAP-PEN}a,d.
The other plots in Figure~\ref{f:freq_NAP-PEN} quantify the accuracy of the MMD method in terms of signed deviation and relative differences of the vibrational frequencies for naphthalene (panels b, c) and pentacene (panels e, f).
For both systems, the approximated frequencies obtained with the MMD method achieve an excellent accuracy for low-frequency modes, corresponding to the shaded regions in Figure~\ref{f:freq_NAP-PEN}b,c,e,f.
In this spectral window, including a number of modes equals to that of the molecular displacements used for building the dynamical matrix, the absolute error is below \SI{0.3}{\per\cm}, corresponding to a relative error of less than \SI{0.2}{\percent}.
The error is hence negligible for the comparison with experimental data as well as for any practical purpose.

The impact of the approximation becomes appreciable for modes at higher energy, shown in the unshaded regions of Figure~\ref{f:freq_NAP-PEN}b,c,e,f.
The MMD frequencies can both overestimate or underestimate the reference values, with maximum absolute deviations of about \SI{20}{\per\cm}, typically for modes above \SI{1000}{\per\cm}.
The relative errors are within \SI{3}{\percent} in magnitude for all the modes.
We note that such a difference is comparable to, if not smaller than, the typical error associated with the semilocal PBE functional in this spectral window, as compared to hybrid density functionals or experiments.
The origin of the discrepancy between MMD and reference calculations is rooted in the approximated V$_\textrm{H}$-V$_\textrm{H}$ block of the dynamical matrix.
We recall that, in the MMD approximation, the V$_\textrm{H}$-V$_\textrm{H}$ block is solely constructed using data from isolated-molecule normal modes calculations, thus disregarding solid-state packing effects on diagonal matrix elements.
In addition, the off-diagonal elements of the V$_\textrm{H}$-V$_\textrm{H}$ block are set to zero, from which we expect a systematic underestimation of the, usually small, phonon bands dispersion of high-frequency modes.
Among other possible sources of discrepancy between MMD and reference frequencies in Figure~\ref{f:freq_NAP-PEN}, it is worth mentioning  the different basis sets employed in the molecular and solid-state calculation, namely Gaussian functions vs.\ plane waves.

We close this section by commenting on the eigenvectors obtained from $\Gamma$ point calculations, on the basis of the overlap matrices between the reference and the MMD  eigenvectors -- see SI Figures~\ref{SI-f:overlap-matrix1}, \ref{SI-f:overlap-matrix2}, \ref{SI-f:overlap}.
This analysis demonstrate a neat one-to-one match between MMD and reference modes in the low-frequency region, corresponding to an overlap matrix is nearly equal to the identity.
For high-frequency modes, the match between reference and approximate eigenvectors holds to a very good approximation, although MMD might occasionally lead to occasional frequency swaps or mode remixing.

\subsection{Phonon dispersion}

\begin{figure}
\includegraphics[width=0.75\textwidth]{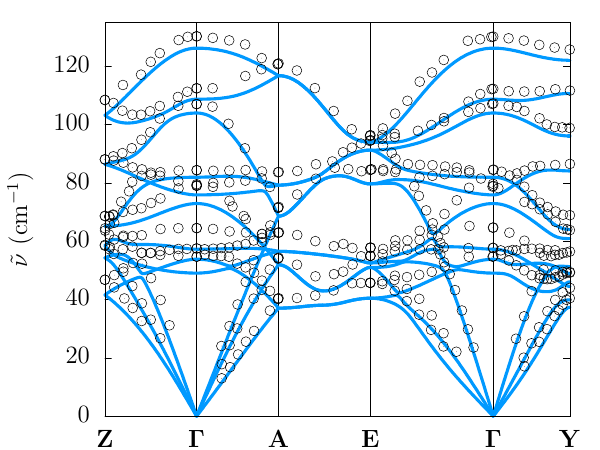}
\caption{Comparison between experimental\cite{Natkaniec80} and simulated (PBE functional with  D3-BJ dispersion corrections)
phonons bands for deuterated naphthalene.
The DFT bands have been obtained by considering a complete set of displacements (reference method). 
}
\label{f:d8-bands}
\end{figure}

We now consider lattice dynamics calculations of molecular crystals fully accounting for phonon bands dispersion. 
Before addressing the comparison between MMD phonons band structures and reference calculations, we confront the latter with experimental data. 
To the best of our knowledge, the only molecular crystal for which phonon bands have been measured with inelastic neutron scattering is fully deuterated naphthalene.\cite{Natkaniec80}
Experimental and calculated phonon bands are superimposed in Figure~\ref{f:d8-bands}. 
Similar to what reported by Kamencek \emph{et al.} \cite{Kamencek20}, an excellent overall agreement is obtained with the PBE functional and D3-BJ dispersion corrections (see Computational details), with a typical accuracy of $\sim$\SI{5}{\per\cm} along the sampled paths in the Brillouin zone. 
Such a favorable agreement attests the adequateness of the DFT level of theory employed in this work as a reference as for it concerns phonons in the THz region.

\begin{figure}
\includegraphics[width=0.85\textwidth]{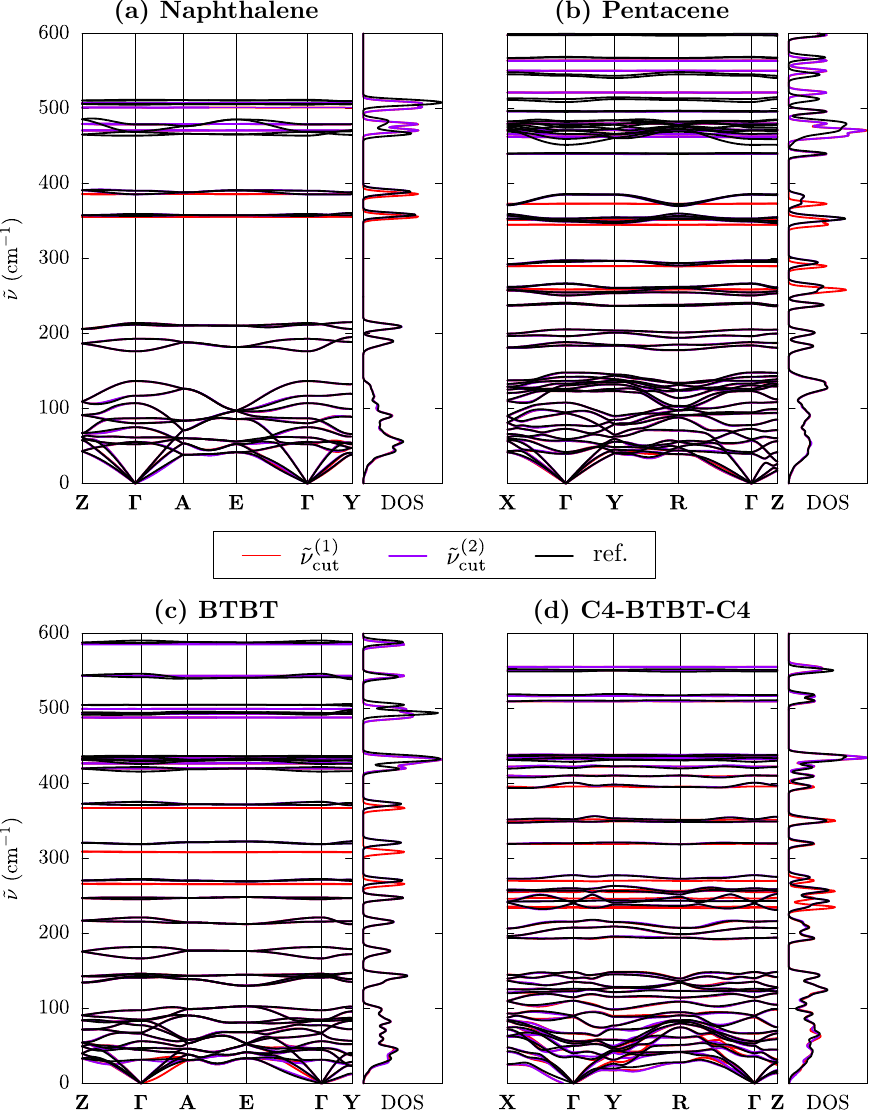}
\caption{Phonon bands and densities of states (DOS) for (a) naphthalene, (b) pentacene, (c) BTBT and (d) C4-BTBT-C4.
Each panel shows the reference and MMD-approximated results obtained for two different cutoff frequencies.
In general, the MMD scheme provides excellent results at wavenumbers up to $\tilde\nu_\mathrm{cut}$, and a good approximation at higher frequencies.
We recall that $\tilde{\nu}_\textrm{cut}^{(1)} \approx \SI{200}{\per\cm}$ and $\tilde{\nu}_\textrm{cut}^{(2)} \approx \SI{400}{\per\cm}$  (see Table~\ref{t:cutoff} for exact cutoff values).
See SI Table~\ref{SI-t:BZ-points} for the definition of high symmetry points in Brillouin zone. }
\label{f:bands}
\end{figure}

Figure~\ref{f:bands} shows phonon bands and densities of states (DOS) for the four systems under scrutiny, comparing MMD-approximated results with reference ones (see SI Figure~\ref{SI-f:bands_hinu} for plots covering the higher-frequency region).
The band structure of the selected crystals presents the typical features of molecular crystals, namely highly dispersive bands in the THz region, leading to a continuum of states, and almost flat bands at higher frequencies.
The latter, giving rise to sharp peaks in the DOS, correspond to dispersionless, Einstein-type, vibrations that closely resemble the modes of individual molecules.

We emphasize that the approximated MMD scheme provides a very accurate description of the low-frequency and highly dispersive region, with only minor differences with respect to the reference band structures and DOSs that are barely visible on the figure scale.
Appreciable discrepancies occur at wavenumbers comparable to or higher than the chosen cutoff, whose magnitude is in line with what observed for $\Gamma$ point modes.
The MMD-approximation exhibits a systematic underestimation of the, however small, phonon bandwidth, resulting in sharper peaks in the DOS.
This is expected because of the neglect of a direct coupling between high-frequency intramolecular displacements (off-diagonal elements of the V$_\textrm{H}$-V$_\textrm{H}$ block).

The MMD method performs equally well in the low-frequency region of dispersive THz modes for all the four examined systems.
Naphthalene, being a pretty rigid molecule (lowest-energy mode for the isolated molecule at $\tilde\nu_0=\SI{169}{\per\cm}$), is a system that is ideally suited for the MMD treatment and good performances were indeed expected.
On the other hand, the other three molecules are more flexible and potentially more challenging than naphthalene for the MMD framework.
Pentacene ($\tilde\nu_0 = \SI{36}{\per\cm}$) and BTBT ($\tilde\nu_0 = \SI{59}{\per\cm}$) present a more flexible conjugated core, while C4-BTBT-C4 ($\tilde\nu_0 = \SI{12}{\per\cm}$) also includes two floppy alkyl chains.
Nonetheless, the MMD approximation remains a sensible one, as it is able to capture the mixing between rigid-body molecular motions and low-frequency intramolecular modes determined by solid-state interactions.

\begin{figure}
\includegraphics[width=1\textwidth]{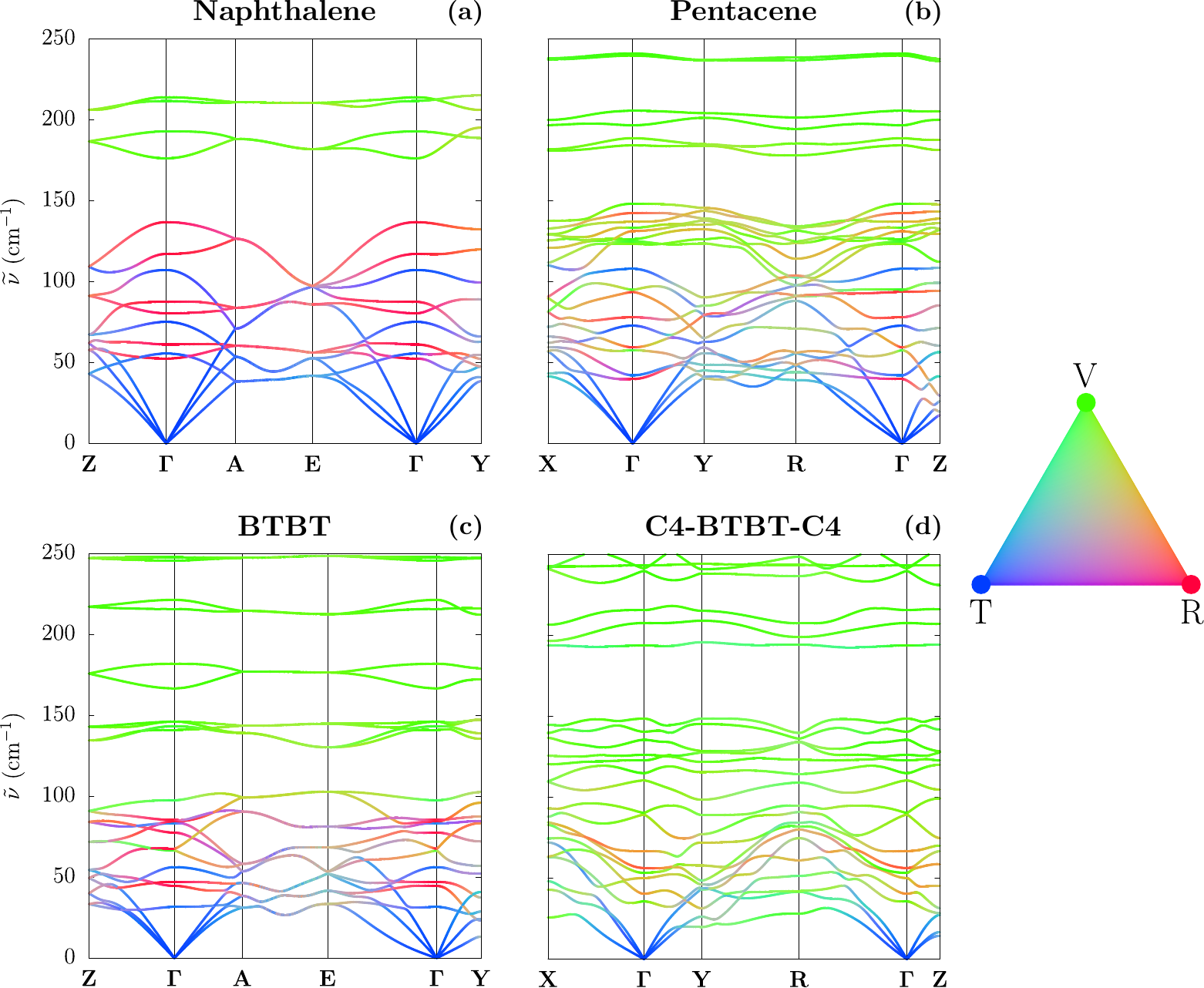}
\caption{Phonon band structures of the four compounds, color-coded according to their translational (T), rotational (R) and intramolecular (V) components, as defined in the ternary color palette.
These images bring a direct insight into the wave vector dependence of the nature of the lattice phonon modes.
}
\label{f:YMCA}
\end{figure}

Having validated the quality of the MMD approach for the calculation of THz modes of molecular crystals, we take advantage of the formulation of the lattice dynamics problem in terms of molecular displacements to gain novel insight on the normal modes.
Figure~\ref{f:YMCA} shows the low-frequency region of the phonon band structures that is color-coded according to the composition of these modes, namely the weight of T, R and V displacements in each eigenvector.

An overview of Figure~\ref{f:YMCA} returns some expected features for the phonon band structure of molecular crystals.
These include the fact that the acoustic branches near $\Gamma$, and more generally the lowest-frequency region of the phonon spectrum, are dominated by translation, whereas high frequency modes are essentially composed of only intramolecular displacements.
More interesting is the intermediate spectral region, from a few tens to about \SI{200}{\per\cm}, in which a significant mixing between displacements of different nature occurs. 
Such a mixing is specific to the different systems under examination and that can be traced back to molecular features and symmetry arguments.

Naphthalene, as anticipated, is a pretty rigid molecule for which a net distinction between rigid-body and intramolecular motion is expected.
This seems supported by the observations of 12 lowest-frequency dispersive bands (i.e.\ 3T+3R per molecule in the unit cell) giving rise to a continuum of vibrational state up to $\sim$\SI{140}{\per\cm}, which is gapped from weakly-dispersive higher-frequency bands.
Indeed, Figure~\ref{f:YMCA}a confirms that the first 12 bands are to a large extent made of T and R displacements.
Consistent with the group factor analysis from rigid molecules, at $\Gamma$-point modes are either pure R or T.
The lower symmetry of Brillouin-zone points at finite $\mathbf{q}$ permits their remixing, leading to the wave vector dependence of the modes nature.

As for it concerns pentacene, BTBT and C4-BTBT-C4 (see Figure~\ref{f:YMCA}b-d), we observe a higher contribution from V displacements in the low-frequency region of the spectrum, due to the higher flexibility of these molecules.
Overall, a large hybridization between T, R and V modes occurs between 30 and \SI{150}{\per\cm}, except at $\Gamma$ where T and R displacements do not mix with each other, but only with V vibrations of appropriate symmetry.
This leads to almost pure R modes near $\Gamma$ or along specific paths for pentacene and BTBT.
Pure or quasi-pure R modes are nearly absent all over the whole Brillouin zone for the highly flexible C4-BTBT-C4 molecule, for which a picture of truly hybrid vibrational modes emerges.

\subsection{Thermodynamic properties}

The calculation of the phonon spectrum grants access to the calculation of the vibrational contribution to thermodynamic properties. 
We hence assess the impact of MMD approximation on the heat capacity on the constant-volume heat capacity, $C_V$ and on the Helmholtz free energy, $\mathcal{F}$, both calculated within the standard statistics of quantum harmonic oscillators as
\begin{eqnarray}
\label{e:Cv}
C_V(T)&=&\frac{k_{\textrm{B}}}{N_q}  \sum_{i=(\mathbf{q},\lambda)} 
\left( \frac{\hbar \, \omega_i}{k_{\textrm{B}} \, T} \right)^2 
\frac{e^{-{\hbar \omega_i}/{k_{\textrm{B}} T}}}
{\left(1-e^{-{\hbar \omega_i}/{k_{\textrm{B}} T}}\right)^2} \\
\label{e:freeene}
\mathcal{F}(T)&=&\frac{1}{N_q} \sum_{i=(\mathbf{q},\lambda)}
\frac{\hbar \, \omega_i }{2} +
k_{\mathrm{B}} T 
\ln{ \left( 1-e^{-{\hbar \omega_i}/{k_{\textrm{B}} T}}\right)}
\end{eqnarray}
where $T$ is the absolute temperature and $k_{\textrm{B}}$ the Boltzmann constant. 
The sums extend over phonon branches and over a grid of $\mathbf{q}$ points sampling the first Brilluoin zone, being $N_q$ the number of samples.

\begin{figure}
\includegraphics[width=0.8\textwidth]{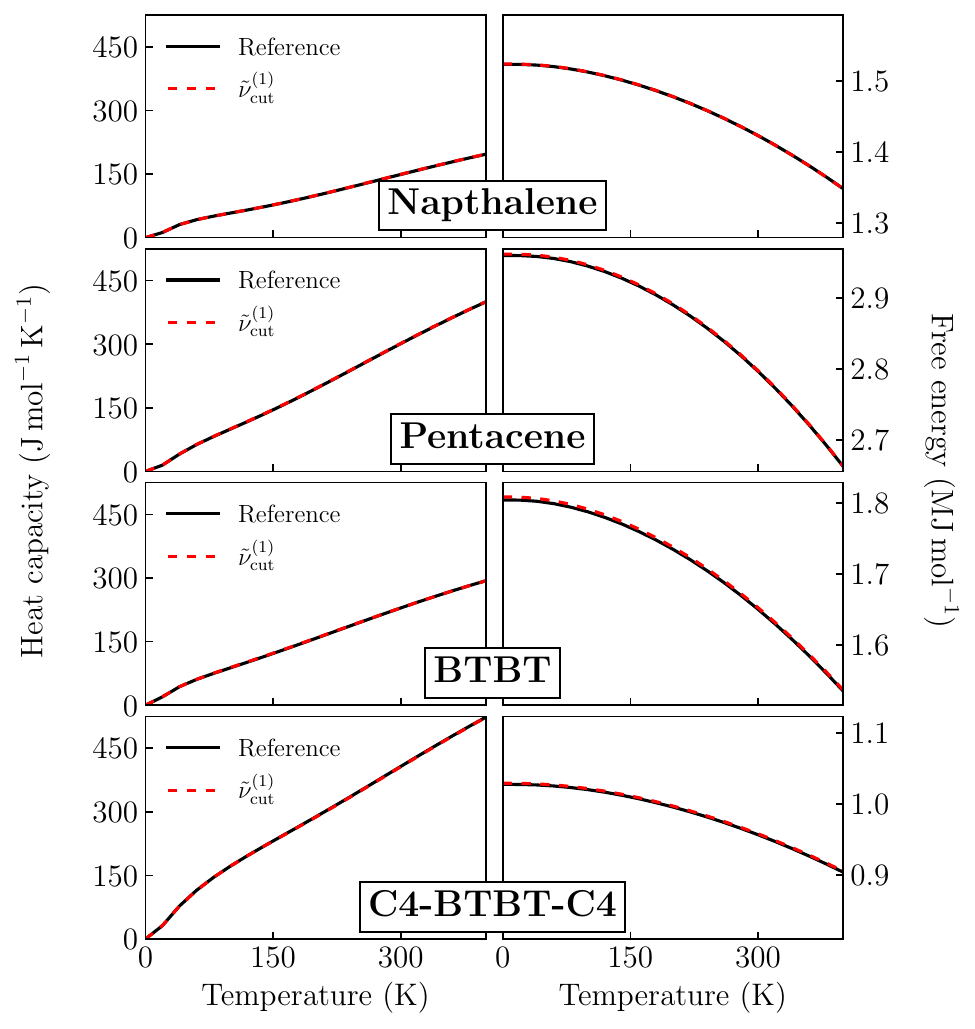}
\caption{Heat capacity (left) and free energy (right) as a function of temperature for the four investigated systems.
An excellent agreement between reference and MMD data is achieved with the lowest cutoff frequency, $\tilde \nu_\mathrm{cut}^{(1)}$.}
\label{f:thermo}
\end{figure}

Figure~\ref{f:thermo} shows the heat capacity and the free energy as a function of $T$ for the four molecular crystals, comparing reference calculations with those obtained within the MMD approximation.
Even adopting a low value for the cutoff frequencies of about \SI{200}{\per\cm}, the MMD approximation achieves an excellent agreement with reference data for both  $C_V$ and $\mathcal{F}$.
The accord on the heat capacity was somehow expected, given the prominent contribution of modes with frequency comparable or lower than thermal energy that are accurately captured within the MMD scheme.

The excellent performances of the MMD approximation on the free energy are less obvious, considering the $T$-independent contribution from zero point motion that is mostly determined by high-frequency modes, which are intrinsically approximated.
The discrepancy between MMD-approximated and reference calculations in Figure~\ref{f:thermo} is of the order of \SI{1}{\kilo\joule\per\mol} in the four system, which corresponds to the typical accuracy required on (free) energy in order to discriminate between crystal polymorphs.\cite{DellaValle08,Zugec24}
Such an agreement can be rationalized with the absence of a systematic bias on the MMD frequencies as compared to reference ones (see Figure~\ref{f:freq_NAP-PEN}), which makes zero-mean errors on phonon frequencies loosely relevant for the result of the sum in Equation~\ref{e:freeene}.
Similarly, the possible systematic underestimation of the band dispersion of high-frequency modes, due to the neglected direct coupling between the displacements of the V$_{\textrm{H}}$-V$_{\textrm{H}}$ block, would have a minimal impact on the free energy.

\subsection{Computational speedup}
\label{s:savings}

The MMD approach allows to largely reduce the number of DFT single point calculations needed to obtain the FC matrix.
Within a conventional central finite differences scheme, and ignoring  space-group symmetry for now (see below), each atom in the unit cell should be displaced back and forth along all three Cartesian directions, which means $6N$ single point calculations, where $N$ is the number of atoms in the unit cell.
On the other hand, in the MMD approach each molecules in the unit cell has to be displaced (twice) along the 3 rigid translations, the 3 rigid rotations and a small number of intramolecular displacements ($N_\textrm{VL}$) obtained by a preliminary calculation on the single molecule -- see Table~\ref{t:cutoff} for typical $N_\textrm{VL}$ values.
Therefore, in the MMD scheme the total number of displacements is $2 Z (6 + N_\textrm{VL} )$, where $Z$ is the number of molecules in the unit cell.
Note that this formula applies to crystals of a single chemical species, the generalization to co-crystals being straightforward.
Knowing $N$ and $N_\mathrm{VL}$, one can readily compute the number of single-point calculations at displaced geometries to be performed in the two approaches and the speedup obtained with the MMD scheme as the ratio between these two numbers, i.e.\
$3N_\mathrm{a}/(6+N_\mathrm{VL})$, where $N_\mathrm{a}=N/Z$ is the number of atoms per molecule.
The value of $N_\mathrm{VL}$ depends on the chosen cutoff (see Table~\ref{t:cutoff}), and for $\tilde \nu_\mathrm{cut}^{(1)}$ we obtain speedups between  6.0 (BTBT) and 7.7 (pentacene), as reported in \hyperref[t:speedup]{\tablename~\ref{t:speedup}}.

\begin{figure}
\includegraphics[width=0.95\textwidth]{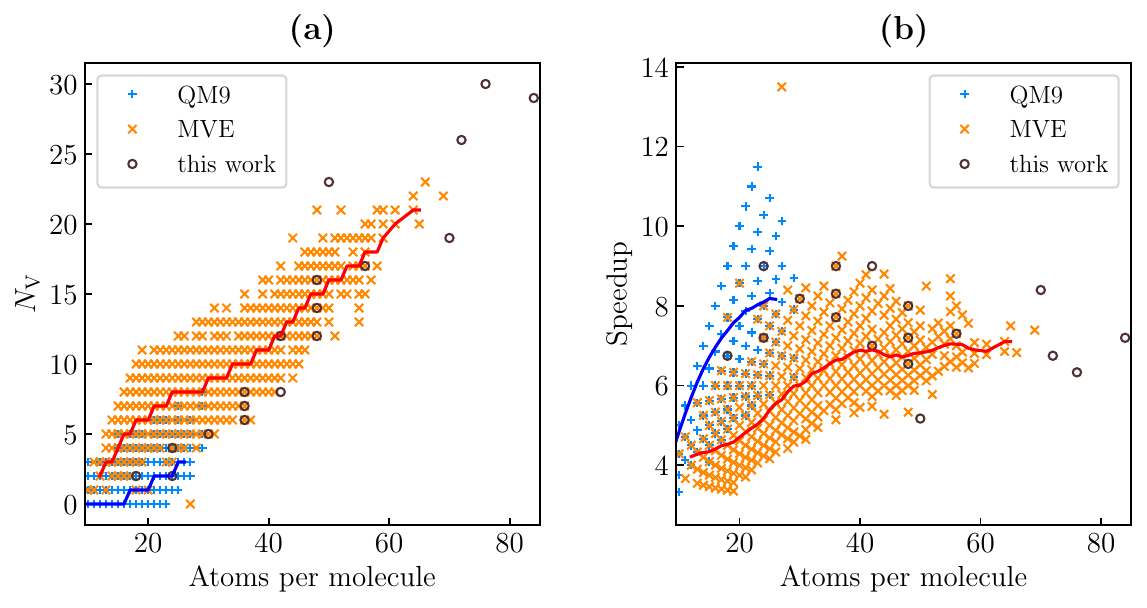}
\caption{(a) Number of single molecule vibrational modes ($N_{\textrm{VL}}$) below \SI{200}{\per\cm} and (b) the corresponding expected speedup for the molecules included in the QM9 \cite{Williams24} and in the MVE-thiol \cite{Koczor22} database and for a set  of  organic semiconductor molecules computed in this work (see SI Table~\ref{SI-t:systems-list}). 
For the two datasets, mobile averages are shown as solid lines.
}
\label{f:speedup}
\end{figure}

In order to give an upper and lower bound for the speedup that can be achieved with the MMD method, we extend the present analysis to a large number of molecular compounds. 
Indeed, an estimate of the speedup can be given on the sole basis of isolated molecule frequency calculations. 
To such an aim, we resort to two published databases of molecular vibrations, namely  QM9 set \cite{Williams24,QM9_DB} (41645 compounds containing H, C, N, O, computed at $\omega$B97x/6-31G* level in vacuum) and the ``Molecular Vibration Explorer" set (MVE, 1907 thiol compounds computed at B3LYP+D3/def2-SVP level in vacuum).
We complemented the two dataset with 18 conjugated molecules employed in organic electronics, including the four compounds studied in this work (see SI Table~\ref{SI-t:systems-list} for the list of compounds), for which we performed  PBE/def2-SVP frequency calculations in vacuum.
For all these molecules, we counted the number of vibrations with frequency $\le \SI{200}{\per\cm}$ and calculated the expected speedup, both shown in Figure~\ref{f:speedup}. 
This allows to conclude that the expected speedup ranges between 4 and 10, with an average value of $\sim 7$ for the compounds above 30 atoms.

\begin{table}[]
\centering
\begin{tabular}{l  S[table-format=3.0]S[table-format=2.0]S[table-format=1.1] S[table-format=3.0]S[table-format=2.0]S[table-format=1.1]}
\toprule
\multirow{2}{*}{System} & \multicolumn{3}{c}{No symmetry} & \multicolumn{3}{c}{Symmetry} \\
            & {AD} & {MMD} & {speedup} & {AD} & {MMD} & {speedup} \\
\midrule
Naphthalene & 216 & 32 & 6.8  &  54 & 11 & 4.9 \\
Pentacene   & 432 & 56 & 7.7  & 216 & 40 & 5.4 \\
BTBT        & 288 & 48 & 6.0  &  72 & 17 & 4.2 \\
C4-BTBT-C4  & 288 & 42 & 6.9  & 144 & 31 & 4.6 \\
\bottomrule
\end{tabular}
\caption{Number of solid-state single point force calculations to be performed for constructing the dynamical matrix with the ordinary atomic displacements (AD) approach and with the MMD method  for $\tilde \nu_\mathrm{cut}^{(1)}$.
The speedup is the ratio between the number of calculations to be performed with AD and that with the MMD scheme.
Results reported neglecting (left-hand columns) and accounting (right-hand columns) for the computational savings due to the exploitation of the point group symmetry.
}
\label{t:speedup}
\end{table}

The crystal symmetry allows to largely reduce the computational workload by avoiding unnecessary calculations for symmetry-equivalent displaced geometries.
As reported in \hyperref[t:speedup]{\tablename~\ref{t:speedup}}, with atomic displacements, the number of calculations can be reduced by a factor that equals the number of symmetry elements belonging to the space group. 
The gain obtained from symmetry is thus more important for the monoclinic crystals (naphthalene and BTBT) with 4 equivalent Wyckoff positions  than for the triclinic systems (pentacene and C4-BTBT-C4) with only 2  equivalent positions.

Also within the MMD scheme, symmetry implies a conspicuous reduction of the number of calculations to be performed.
The gain, now ranging between 4.2 (BTBT) and 5.4 (pentacene), is about 30\% smaller than what obtained with atomic displacements -- see Table~\ref{t:speedup}, meaning that the largest speedups are achieved for low-symmetry structures.
This is due to the fact that collective molecular displacements are less likely than atomic ones to be transformed into an equivalent coordinate by the application of a symmetry operation.
In particular, symmetry elements bringing one molecule into an equivalent one, do systematically reduce the calculations to be performed. 
This is not always the case for symmetry operations connecting different atoms belonging to the same molecule. 
This is partially compensated by the fact that positive and negative rigid-body displacements are equivalent by symmetry in the common case of molecules whose center of mass lays on a symmetry element.
The latter argument holds true also for atomic displacements, but the reduction of the number of displacements is way less important.

Overall, the full implementation of space-group symmetry operations largely reduces the computational cost of phonon calculations in the MMD displacement. 
By taking a pentacene $2\times2\times2$ supercell (576 atoms) as an example, each single point calculation (self consistent field convergence and forces) at a displaced geometry requires approximately \SI{650}{\s}, running in parallel on 384 cores on  8 Intel Skylake compute nodes at \SI{2.7}{\giga\Hz} frequency.
This value refers to DFT calculations starting from the previously converged electronic density of the equilibrium structure.
This corresponds to a computational cost for building the force constant matrix (considering symmetry)
of  $\sim$15000 core hours in the standard approach, which reduces to  $\sim$2800  core hours  with the MMD scheme for $\tilde \nu_\mathrm{cut}^{(1)}$.

\section{Discussion and Conclusions}
\label{s:conclusions}    

In this work, we have presented an original formulation of harmonic lattice dynamics of molecular crystals based on molecular displacements in a frozen phonon framework.
This approach sets a natural scheme for introducing a minimal molecular displacement (MMD) approximation, which permits to obtain an accurate calculation of the vibrational properties, comparable to state of the art solid-state DFT calculations, at a fraction of the computational cost.
In the low-frequency region of thermal phonons, for which our approximated framework is specifically tailored, the deviation between MMD and reference phonon frequencies, eigenvectors and bands turned out to be practically negligible.
The MMD method allows also computing the vibrational contribution to thermodynamic quantities with excellent accuracy.
Besides performance considerations, it is worth adding that the molecular formulation of the vibrational problem offers a direct molecular-level insight onto the complex lattice dynamics of molecular crystals.

The present development employs isolated-molecule normal mode calculations to set up a basis of intramolecular displacements for solid-state calculations and to complement the high-frequency sector of the dynamical matrix.
This approach demonstrates excellent performances for the compounds selected in this study, albeit molecules with multiple stable conformers or with floppy moieties might pose additional challenges, as in those cases the equilibrium geometry of the molecules can significantly differ between the gas and the solid state.
In this sense, the quality of the results obtained for C4-BTBT-C4, similar to that of the other more rigid molecules considered in this study, seems to confirm the robustness of the adopted methodology, even for a compound presenting flexible alkyl chains. 
As a possible improvement, we mention that the isolated-molecule analysis can be replaced by QM/MM calculations for single molecules (QM subsystem) in their  crystalline MM environment.
This would fix issues associated with the molecular equilibrium geometry without affecting the computational cost.
We conjecture that a QM/MM treatment of intramolecular modes could yield to reliable phonon calculations for smaller cutoff frequencies, thus further lowering the global computational cost.
Moreover, QM/MM molecular calculations could improve the agreement with reference data in the high frequency region. 

In this work, devoted to the presentation and benchmarking of the MMD approximation, molecular and solid-state calculations have been all run with the same semilocal functional PBE for internal consistency.
However, different choices can be made, such as employing a hybrid functional for the  molecular calculations, taking advantage of a superior description of vibrations at a modest increase of the computational cost.
We emphasize that such a mixed functional approach has the potential to improve the description of the high-frequency region of the vibrational spectrum as compared to experiments.

The MMD method paves the way toward the accurate and cost-effective large-scale computational endeavors, such as a wide \emph{in~silico} screening of the vibrational properties of organic solids or the production of accurate reference data sets for the training of machine learning potentials for molecular crystals.
The MMD scheme also offers a robust and convenient platform for computing infrared and Raman intensities or for introducing anharmonic effects.
State-of-the-art anharmonic phonon procedures such as those implemented in SSCHA\cite{SSCHA} or SCAILD/QSCAILD \cite{SCAILD,QSCAILD} are straightforwardly compatible with our formalism, as it allows for the treatment of anharmonic effects only in the reduced subspace. An exploration of this will be subject of future work.
We also emphasize that the frequency-wise divide-and-conquer strategy proposed in this paper could be generalized to the computation of phonons in covalent solids,  alleviating the computational burden in systems with large unit cells, e.g.\ in covalent or metal organic frameworks or hybrid perovskites. 
Finally, we foresee the application of lattice dynamics with the MMD scheme, in combination with electron-phonon coupling calculations, to the accurate, efficient and insightful study of the dynamic energetic disorder in high-mobility organic semiconductors.

\begin{acknowledgement}
GD also thanks Alberto Girlando for useful correspondence.
This work received financial support from the French ”Agence Nationale de la Recherche”, project RAPTORS (ANR-21-CE24-0004-01).
This work was performed using HPC resources from GENCI-TGCC (Grant No.\ 2021-A0110910016)
\end{acknowledgement}

\begin{suppinfo}
This will usually read something like: ``Experimental procedures and
characterization data for all new compounds. The class will
automatically add a sentence pointing to the information on-line:
\end{suppinfo}

\bibliography{phonons}

\end{document}